\documentclass[11pt]{article}

\usepackage{amssymb}
\usepackage{amsmath}
\usepackage{latexsym}

\newcommand{\beq}{\begin{equation}}
\newcommand{\eeq}{\end{equation}}

\def\bb {\begin {eqnarray}}
\def\ee {\end {eqnarray}}

\newcommand{\bea}{\begin{eqnarray}}
\newcommand{\ena}{\end{eqnarray}}

\newcommand{\beano}{\begin{eqnarray*}}
\newcommand{\enano}{\end{eqnarray*}}

\newcommand{\bee}{\begin{enumerate}}
\newcommand{\ene}{\end{enumerate}}

\newcommand{\bei}{\begin{itemize}}
\newcommand{\eni}{\end{itemize}}
\newcommand\ei{\end{itemize}}

\newcommand{\bs}{\begin{slide}}
\newcommand{\es}{\end{slide}}

\newcommand{\bra}{\begin{array}}
\newcommand{\era}{\end{array}}
\newcommand{\bqn}{\begin{eqnarray}}
\newcommand{\eqn}{\end{eqnarray}}
\newcommand\ben{\begin{enumerate}}
\newcommand\een{\end{enumerate}}

\def\BC{\mathbb C}
\def\_\BC{\mathbbi C}

\def\K{{\mathcal K }}
\def\W{{\mathcal W }}
\def\H{\mathbb{H}}
\def\R{\mathbb{R}}

\def\C{\mathbb{C}}

\def\de Sitter {de Sitter}

\newcommand{\be}{\beta}

\def\H{{\cal H}}
\def\K{{\cal K}}

\newcommand{\lbl}[1]{\label{eq: #1}}

\def\R{{\rm I\hspace{-.15em}R}}

\def\C{\hspace{3pt}{\rm l\hspace{-.47em}C}}

\def\b{\begin{equation}}
\def\e{\end{equation}}
\def\bd{\begin{displaystyle}}
\def\ed{\end{displaystyle}}
\def\ba{\begin{array}}
\def\ea{\end{array}}

\def\bee{\begin{enumerate}}
\def\eee{\end{enumerate}}

\def\bes{\begin{eqnarray*}}
\def\ees{\end{eqnarray*}}
\def\be{\begin{eqnarray}}
\def\ee{\end{eqnarray}}

\usepackage{a4wide}

\begin{document}
\title{A discrete nonetheless remarkable brick in de Sitter: \\
  the ``massless minimally coupled field''}


\date{}

\author{Jean Pierre Gazeau and Ahmed Youssef
\footnote{e-mail: gazeau@apc.univ-paris7.fr,  youssef@apc.univ-paris7.fr}\\
\\
\emph{Laboratoire APC,
Universit\'e Paris 7-Denis Diderot}}

\maketitle

\begin{abstract}
Over the last ten years interest in the physics of de Sitter spacetime has been growing very fast. Besides the supposed existence of a ``de sitterian period" in inflation theories, the observational evidence of an  acceleration of the universe expansion (interpreted as a positive cosmological constant or a ``dark energy'' or some  form of ``quintessence'')  has triggered a lot of attention in the physics community. A specific de sitterian field called ``massless minimally coupled  field" (mmc) plays a fundamental role in inflation models and in the construction of the de sitterian gravitational field. A covariant quantization of the mmc field, \emph{\`a la} Krein-Gupta-Bleuler was proposed in \cite{gareta}. In this talk, we will review this construction and explain the relevance of such a field in the construction of a massless  spin 2 field in de Sitter space-time.
\end{abstract}

\section{Introduction}\label{int1}
Let us begin with this citation from the well known book by Birrell and Davies \cite{BD} 
\begin{quote}
	``The quantization of the gravitational field has been pursued with great ingenuity and vigor over the past forty years, but a complete satisfactory quantum theory of gravity remains elusive.''
\end{quote}
But what about the situation almost three decades after this statement? Although many programs aiming to construct a consistent quantum gravity theory, such as string theory \cite{pol}, loop quantum gravity \cite{rov}, causal sets \cite{sor}, dynamical triangulations \cite{loll} and asymptotic safety \cite{reni} have been developed, none can claim to have attained yet a reasonable level of maturity enabling computations in physically relevant situations. In this context, all one can do is to treat gravity as a low energy effective field theory \cite{burg}. This is the rationale behind perturbative quantum gravity around fixed backgrounds. 

The first step toward this goal is to construct a consistent linear quantum theory of gravity. Aside from the usual conceptual difficulties encountered in any attempt to quantize gravity, this program has met reasonable success when applied around Minkowski spacetime without raising significant technical difficulties. However, as soon as one tries to construct linear quantum gravity around a non trivial background, one faces more serious technical and conceptual difficulties. Due to its maximal symmetry and to its relevance to cosmology, one of the  most natural backgrounds  one is interested in is the de Sitter (dS) spacetime, and no wonder that a huge literature was devoted to the study of QFT and perturbative quantum gravity on de Sitter space (see the excellent review \cite{antorev} and references therein).

As we will see in this paper, the principle of covariance, which plays such  a fundamental role in flat spacetime, is delicate to implement in any gauge theory in dS space, including linear quantum gravity   \cite{antorev,woodbreak}. The author of \cite{woodbreak} talks about a fundamental inconsistency ``when combining the causal properties of de Sitter and the constraint equations of any gauge theory  [because] a de Sitter invariant gauge converts elliptic constraint equations into hyperbolic evolution equations''. More precisely, if one insists on dS covariance - for instance by working on $S^4$, the Euclidean version of the de Sitter space and then analytically continuing the solution to the Lorentzian one- one obtains a pathological graviton propagator which grows  unboundedly both at large spacelike and large timelike separations \cite{altu} (see however  \cite{hertog} and \cite{higuchi}). If one abandons covariant methods like the Euclidean continuation and quantizes the graviton by canonical methods, the dS covariance is very hard to maintain \cite{antorev}. Similar problems arise in the quantization of the mmc scalar field on de Sitter. 

In this talk, we will first prove that the graviton field can be built in terms of a mmc scalar field on de Sitter. We then present the quantization of the graviton field \emph{\`a la} Wightman \cite{stwi} based on the existence of a well behaved two point function. Our work has a strong group theoretical content, which  turns out being a convenient framework to address the non trivial question of gauge fixing arising on curved backgrounds. This gauge fixing problem is at the heart of the difficulties encountered earlier in the construction of the graviton propagator on de Sitter \cite{gravitonprb}. We wish to make clear in a future work \cite{gayou} that the bad long distance behaviour of the graviton propagator on dS is a gauge artifact that can be circumvented in our approach. Finally we review a quantization method \emph{\`a la} Krein-Gupta-Bleuler proposed in \cite{gareta} that yields a covariant quantization of the mmc field and that might be relevant to the quantization of the graviton field.

\section{De Sitter linear quantum gravity: the field equations}\label{sec1}
In this work, we  restrict ourselves to linear perturbations of Einstein gravity around the de Sitter metric (gravitational waves in de Sitter spacetime) and their quantization, neglecting the graviton-graviton interactions. The counting of dynamical  degrees of freedom of the theory goes as follows: due to the parametrization invariance, only $6$ of the $10$ independent components of the symmetric metric tensor $g_{\mu \nu}$ are physical degrees of freedom. Then the Bianchi identity $\nabla^\mu G_{\mu \nu}=0$ - where $G_{\mu \nu}=R_{\mu \nu}-\dfrac{1}{2} R g_{\mu \nu}$ is the Einstein tensor - shows that 4 equations of motion are constraints and not dynamical equations. We finally have that only $2$ degrees of freedom are propagating. In order to study linearized gravity in de Sitter spacetime, we begin by splitting our metric in a de Sitter fixed background $\hat{g}_{\mu \nu}$ and a small fluctuation $h_{\mu \nu}$ such that $g_{\mu \nu} = \hat{g}_{\mu \nu} + h_{\mu \nu}$. The reparametrization invariance is translated at the linear level as the gauge invariance:
\b
\label{gaugeX}
h_{\mu\nu} \longrightarrow h_{\mu\nu}+\nabla_{\mu}\Xi_{\nu}+\nabla_{\nu}\Xi_{\mu}\, ,
\e
where $\Xi_{\nu}$ is an arbitrary vector field.
The linearized Lagrangian contains no term proportional to $h^2$, hence the usual description of the graviton field as a \emph{massless} spin 2 field. It is important to realize at this point that  the concept of mass/masslessness is ambiguous in a generic curved spacetime since it relies heavily on the existence of a flat background metric. Note however that the mass concept can be made precise in the context of (anti-)de Sitter space thanks to the maximal symmetry of these spaces \cite{gano,flfrga}. The wave equation for ``massless'' tensor fields $h_{\mu\nu}$ propagating in de Sitter space takes the form 
\begin{equation}
\label{equationX}
(\Box+2H^2)h_{\mu\nu}-(\Box-H^2) \hat{g}_{\mu\nu}h'-\left(\nabla_\mu \nabla^{\rho}h_{\nu\rho}+\nabla_\nu \nabla^{\rho}h_{\mu\rho}\right)
+\hat{g}_{\mu\nu}\nabla^{\lambda}\nabla^{\rho}h_{\lambda\rho}
+\nabla_{\mu}\nabla_{\nu}h'=0\, .
\end{equation}
Here, $H$ is the Hubble constant, $\nabla^\nu$  is the covariant derivative on de Sitter space, $\Box=\hat{g}_{\mu \nu} \nabla^\nu \nabla^\mu$ is the Laplace-Beltrami operator and $h'=\hat{g}^{\mu \nu} h_{\mu \nu}$.  At this level one can choose either to fix the gauge, or to work only with gauge invariant variables, the Bardeen variables, which can be found in closed form in the linearized theory \cite{bardeen,stewart}. In the following we will partially fix the gauge by adding to the Lagrangian a gauge fixing term that is a generalization of the Lorentz/harmonic gauge condition 
\b
\label{gaugecondX}
\nabla^\mu h_{\mu \nu}= \xi \nabla_\nu h'\, .
\e 
We will show in the next section that the special value $\xi=1/2$ makes the relation between the field and group representation thoroughly apparent. 
\section{De Sitter group interpretation}\label{sec2}
Our aim is to construct de Sitter elementary systems (in the Wigner sense) in analogy with the Minkowskian case. Recall that in the minkowskian case, the field equations are the Casimir eigenvalues equations with eigenvalues  $m^2$ and $s$, the mass and the spin. In fact the latter carry the group-theoretical content of the theory. We thus wish to describe the de Sitter tensor field equation ($\ref{equationX}$) as an eigenvalue equation of a $SO_0(1,4)$ Casimir operator (the subscript $0$ stands for the subgroup of $SO(1,4)$ connected to the identity). It will be convenient to use the ambient space coordinates defined hereafter in order to express the spin-2 field equation in terms of the coordinate independent Casimir operator.  
In the following, we describe the de Sitter space as a one-sheeted hyperboloid embedded in a five-dimensional Minkowski space. This choice of coordinates makes manifest the $SO_0(4,1)$ invariance:
$$
M_{dS}\equiv \left\{x \in \R^5; x^2= \eta_{\alpha \beta} x^\alpha x^\beta=-H^{-2}\right\}\, ,
$$
where $\eta_{\alpha \beta}=\textrm{diag}(1,-1,-1,-1,-1)$. The de Sitter metric is the induced metric on the hyperboloid
$$
ds^2=\eta_{\alpha \beta} dx^\alpha dx^\beta \left.\right|_{x^2=-H^{-2}}=\hat{g}_{\mu \nu} dX^\mu dX^\nu, \mu=0,1,2,3\, ,  
$$
where the $X^\mu$'s are intrinsic spacetime coordinates for the de Sitter hyperboloid.
In ambient space notations, the tensor field $\K_{\alpha\beta}(x)$ can be viewed as a homogeneous function in the $\R^5$ variables $x^\alpha$ with some arbitrarily chosen degree $\sigma$ :
$$ x^{\alpha}\frac{\partial }{\partial
x^{\alpha}}\K_{\beta\gamma}(x)=x\cdot\partial \K_{\beta\gamma}
(x)=\sigma \K_{\beta\gamma}(x)\,.$$
Not every homogeneous field of $\R^5$ represents a physical dS entity. In order to ensure that $\K$ lies in the dS tangent spacetime, it must also satisfy the transversality condition 
$$ x\cdot\K(x)=0,\mbox{ \it i.e. }x^\alpha
\K_{\alpha\beta}(x)=0,\mbox{ and } x^\beta \K_{\alpha\beta}(x)=0\, .
$$
Given the importance of this transversality property of dS fields, let us introduce the symmetric transverse projector  $\theta_{\alpha \beta}=\eta_{\alpha \beta}+H^2x_{\alpha}x_{ \beta}$. It is the transverse form of the dS metric $\hat{g}_{\mu\nu}$:
\b
\hat{g}_{\mu\nu}=\frac{\partial x^{\alpha}}{\partial
X^{\mu}}\frac{\partial x^{\beta}} {\partial
X^{\nu}}\,\theta_{\alpha\beta}\, .
\e
The transverse projector is used in the construction of transverse entities such as the tangential (or transverse) derivative $ \bar
\partial_\alpha=\theta_{\alpha \beta}\partial^\beta=
\partial_\alpha  +H^2x_\alpha x\cdot\partial$\, .

In order to express Eq. (\ref{equationX}) in terms of the ambient coordinates, we use the fact that the ``intrinsic'' field $h_{\mu\nu}(X)$ is
locally determined by the ``transverse'' tensor field $\K_{\alpha\beta}(x)$ through 
\b\lbl{passage}
h_{\mu\nu}(X)=\frac{\partial x^{\alpha}}{\partial
X^{\mu}}\frac{\partial x^{\beta}}{\partial
X^{\nu}}\K_{\alpha\beta}(x(X))\, . 
\e 
In order to establish the relation of the tensor field with the irreducible representation of the de Sitter group we present the field
equation in terms of the second order Casimir operator. A familiar realization of the Lie algebra of the de Sitter group is the one generated by the ten Killing vectors $K_{\alpha \beta}=x_\alpha \partial_\beta -x_\beta \partial_\alpha$.
The second order Casimir operator is defined in terms of the self-adjoint $L_{\alpha \beta}$ representatives of the Killing vectors: 
\b Q_2=-\frac{1}{2}
L^{\alpha\beta}L_{\alpha\beta}=-\frac{1}{2}
 (M^{\alpha\beta}+S^{\alpha\beta})(M_{\alpha\beta}+S_{\alpha\beta})\, ,\e
where $M_{\alpha\beta}=-i (x_\alpha \partial_\beta-x_\beta
\partial_\alpha)=-i (x_\alpha \bar \partial_\beta-x_\beta \bar
\partial_\alpha)$ is the orbital part, and the subscript 2 reminds that the carrier space is constituted by second rank tensors. The action of the spinorial
part $S_{\alpha\beta}$ is given by \cite{gaha} \b
S_{\alpha\beta}\K_{\gamma\delta}=-i
(\eta_{\alpha\gamma}\K_{\beta\delta}-\eta_{\beta\gamma}\K_{\alpha\delta}
+\eta_{\alpha\delta}\K_{\beta\gamma}-\eta_{\beta\delta}\K_{\alpha\gamma})\, .\e
We also define the scalar part of the Casimir operator
$Q_{0}=-{{1}\over{2}}M_{\alpha\beta}M^{\alpha\beta}=-H^{-2}(\bar\partial)^2$.
In terms of the previously defined operators,  the field equation for $\K$ takes the form \cite{fr,gareta}
\begin{equation}
\label{fieldK}
(Q_2+6)\K(x)+D_2\partial_2\cdot\K=0\, .
\end{equation}
where the operator $D_2$ is the generalized gradient $ D_2K=H^{-2}{\cal S}(\bar
\partial-H^2x)K$, and ${\cal S}$ is the symmetrizer operator (${\cal S}\xi_{\alpha}\omega_{\beta}=\xi_{\alpha}\omega_{\beta}+\xi_{\beta}\omega_{\alpha}$). The operator $\partial_2.$ is the generalized divergence on the dS  hyperboloid $
\partial_2\cdot \K=\partial
\cdot\K- H^2 x \K'-\dfrac{1}{2} \bar  \partial \K'$, where $\K'$ is the trace of $\K_{\alpha \beta}$. 
As expected, this formulation of the field equation has now a clear group-theoretical content. 
Using the identities 
\b
\partial_2\cdot
D_2\Lambda=-(Q_1+6)\Lambda \qquad  Q_2D_2\Lambda=D_2Q_1\Lambda\, ,
\e
where the action of the Casimir operator $Q_1$ on a vector field $\Lambda$ is given by \cite{gagata,gata}
\b 
Q_1\Lambda(x)=(Q_0-2)\Lambda(x)+2x_\alpha \partial\cdot
\Lambda-2\partial_\alpha x\cdot \Lambda\, .
\e
one can simply show that the gauge invariance (\ref{gaugeX}) of the field equation is expressed in the ambient coordinates as
\b 
\K \rightarrow \K+D_2\Lambda\, .
\e
and the general family of gauge conditions (\ref{gaugecondX}) in ambient space notations takes the form 
\b
\partial_2\cdot\K=(\zeta-\frac{1}{2})\bar
\partial \K'\, .
\e

The field equation (\ref{fieldK}) can be derived from the following action 
\b\lbl{lagrangien}
 S=\int d\sigma{\cal L}, \;\;
{\cal L}=-\frac{1}{2x^2}\K\cdot \cdot(Q_2+6)\K+\frac{1}{2}(\partial_2
\cdot\K)^2\, ,
\e
where $d\sigma$ is the volume element in dS space. After adding a gauge fixing term to the Lagrangian we get 
\b 
{\cal L}=-\frac{1}{2x^2}\K..(Q_2+6)\K+\frac{1}{2}(\partial_2
\cdot\K)^2+\frac{1}
{2\alpha}(\partial_2\cdot\K-(\zeta-\frac{1}{2})\bar
\partial \K')^2\, .
\e
Hereafter we shall work with the specific gauge $\zeta=\frac{1}{2}$, since it vividly exhibits the relation between the tensor field and the dS group representation. With this choice $(\zeta=\frac{1}{2})$, we obtain 
\b
\label{lagrangianc}
{\cal L}=-\frac{1}{2x^2}\K..(Q_2+6)\K+\frac{c}{2}(\partial_2 \cdot\K)^2\, .
\e
\b
\label{equationc}
(Q_2+6)\K(x)+cD_2\partial_2 \cdot\K=0\, ,
\e 
where $c=\dfrac{1+\alpha}{\alpha}$ is a gauge fixing parameter. In contrast with the flat space case, the simplest choice of $c$ is not zero (Feynman gauge). Finding the optimal value of $c$ is actually a non trivial question on curved backgrounds and have important consequences on the two point function that we will develop in the next section.  
\section{Solutions of the tensor wave equation}
A general solution of Eq. (\ref{equationc}) can be constructed from a combination of a scalar field and two vector fields. Let us first introduce a traceless tensor field $\K$ in terms of a five-dimensional constant vector $Z_1=(Z_{1\alpha})$, a scalar field $\phi_1$ and two vector fields $K$ and $K_g$ by putting 
\b
\label{ansatz}
\K=\theta\phi_1+ {\cal S}\bar
Z_1K+D_2K_g,\;\K'=4\phi+2Z.K+2H^{-2}\bar
\partial\cdot K_g=0\, ,
\e
where $\bar Z_{1\alpha}=\theta_{\alpha\beta}
Z^{\beta}_1,$ and $ x\cdot K=0=x\cdot K_g$. Applying (\ref{equationc}) to the above
ansatz, using the commutation rules and algebraic identities for the various involved operators and fields and the value $c=2/5$ for the gauge fixing parameter, we can construct the tensor field $\K$ in terms of a ``massless'' minimally coupled scalar field $\phi$ (see \cite{gagata,LinCov} for more details)
\b 
\label{tensorscalar}
\K_{\alpha
\beta}(x)=\K^{2/5}_{\alpha
\beta}(x)={\cal D}_{\alpha \beta}(x,\partial,Z_1,Z_2)\phi\, ,\;\;
Q_0 \phi=0\, ,\e 
where  ${\cal D}$ is the projector tensor 

\begin{eqnarray}
{\cal D}(x,\partial,Z_1,Z_2)= && \left(-\frac{2}{3}\theta Z_1\cdot +{\cal S}\bar
Z_1+\frac{1}{9 }D_2 (H^2 x\cdot Z_1-Z_1.\bar \partial+\frac{2}{3}H^2
D_1 Z_1\cdot)\right) \\
\nonumber
&& \left( \bar Z_{2}-\frac{1}{2} D_{1}(Z_2.\bar
\partial+2H^2x\cdot Z_2)\right)\, .
\end{eqnarray}

The two arbitrary constant vector $Z_1$ and $Z_2$ can be fixed in terms of the polarization
tensor of gravitational field in the minkowskian limit.

In order to see in which sense the value $c=2/5$  is the optimal one let us  briefly consider the case $c\neq2/5$ and $c\neq 1$. In this case one can show that the field takes the form \cite{LinCov}
\b
\K^c=\K^{2/5}+\frac{2-5c}{5 (1-c)} D_2 (Q_1+6)^{-1} \partial\cdot\K^{2/5}\, .
\e
The extra term $D_2 (Q_1+6)^{-1} \partial. \K^{2/5}$ is responsible of logarithmic divergences in the field solutions and in the two point function
(see \cite{galmp,ga} and references therein for a detailed review on the gauge fixing problems in AdS space, formally similar to dS space,  and the group structure behind it).

\section{The two-point function}
In a previous paper devoted to the ``massive'' spin-$2$ field
(divergenceless and traceless) \cite{gagata}, the construction of the quantum
field from the Wightman two-point function ${\cal W}$ has been carried out. This
function fulfills the conditions of: a) positiveness, b) locality,
c) covariance, d) normal analyticity, e) transversality, f)
divergencelessness and g) permutational index symmetries.
We have found that this function can be written in the form \b {\cal
W}^\nu_{\alpha\beta \alpha'\beta'}(x,x')=D_{\alpha\beta
\alpha'\beta'}(x,x'){\cal W}^\nu(x,x'), \e where $\nu$ is related to a ``mass'' (understood through the Minkowskian limit) by $m^2=H^2 (\nu^2+\dfrac{9}{4})$ , ${\cal
W}^\nu(x,x')$ is the Wightman two-point function for the massive
scalar field and $D_{\alpha\beta \alpha'\beta'}(x,x')$ is a
projection bi-tensor. We could crudely replace $\nu$
 by $\pm \frac{3}{2}i$ in order to get
the ``massless'' tensor field associated to linear quantum gravity
in dS space. However this naive procedure leads to  singularities in the definition of the Wightman two-point function.

These difficulties in taking the massless limit is a well known feature in most field theories, and is due to the gauge symmetry responsible of the masslessness. It has in fact been proven that the use of an indefinite metric is an unavoidable feature if one insists on preserving the locality and covariance in gauge quantum field theories \cite{strocchi}. The positivity requirement in the Wightman axioms cited above have to be relaxed, and one must construct a quantization \emph{\`a la} Gupta-Bleuler in order to extract in a consistent manner the physical subspace of states on which the inner product will still be positive definite. More precisely in our case we define $V_{c}$ as the space of solutions of $(Q_2+6)\K(x)+cD_2\partial_2 \cdot\K=0$ which are square integrable with respect the (degenerate) dS invariant inner product 
\begin{equation*}
(\K_1,\K_2)=\frac{i}{H^2}\int_{\renewcommand{\arraystretch}{0.4}
\!\!\begin{array}{l}
\begin{scriptstyle}S^3\end{scriptstyle}\\ \begin{scriptstyle}\rho=0
\end{scriptstyle}\end{array}}\!\!\! [\K_{1}^*\cdot \cdot
{\partial}_{\rho} \K_2-c({\partial}_{\rho}
x.\K_{1}^*)\cdot (\partial \cdot \K_2)-(1^*\rightleftharpoons
2)]d\Omega\, .\end{equation*} 
This inner product is defined in terms of bounded global intrinsic
coordinates $(X^\mu,\;\mu=0,1,2,3)$ well-suited to describe a
compactified version of dS space, namely S$^3 \times{\rm S}^1$. There exists a ``Gupta-Bleuler triplet'' $V_g \subset V \subset
V_{c}$ carrying an indecomposable representation of the de Sitter group.
\begin{itemize}
  \item $V$ is the closed subspace of $V_c$ of solutions satisfying
the divergencelessness condition. The inner product is positive semidefinite when restricted to the subspace $V$.
  \item $V_g$ is the subspace  of $V$ consisting of gauge  solutions of the form $\K_g=D_2\Lambda_g$ with a vector field $\Lambda_g$.
These are orthogonal to every element in $ V$ including themselves.
  \item The physical states belong to the quotient space $V/V_g$ where the inner product is positive definite. 
\end{itemize}

Let us briefly recall the required conditions for the ``massless''
bi-tensor two-point function ${\cal W}_{\alpha\beta\alpha'\beta'}(x,x')$, where $x,x'\in X_H$. These functions entirely  encode
 the theory of the generalized free fields on dS
space-time $X_H$. They have to satisfy the following requirements:
\begin{enumerate}
\item[a)] {\bf Indefinite sesquilinear form}
for any test function $f_{\alpha \beta} \in {\cal D}(X_H)$, we
have an indefinite sesquilinear form that is defined by
\begin{equation} \int _{X_H \times X_H} f^{*\alpha\beta}(x)
{\cal W}_{\alpha\beta\alpha'\beta'}(x,x')f^{\alpha'\beta'}
(x')d\sigma(x)d\sigma(x')\, ,\end{equation} where $ f^*$ is the
complex conjugate of $f$ and $d\sigma (x)$ denotes the
dS-invariant measure on $X_H$. ${\cal D}(X_H)$ is the space of
$C^\infty$ functions $X_H \mapsto \C$   with compact support in $X_H$.
\item[b)] {\bf Locality}
for every spacelike separated pair $(x,x')$, {\it i.e.} $x\cdot
x'<H^{-2}$,
\begin{equation}
{\cal W}_{\alpha\beta \alpha'\beta'}(x,x')={\cal
W}_{\alpha'\beta'\alpha\beta }(x',x) \, .\end{equation}
 \item[c)] {\bf Covariance}
     \begin{equation}
(g^{-1})^{\gamma}_{\alpha}(g^{-1})^{\delta}_{\beta} {\cal
W}_{\gamma\delta \gamma'\delta'} (g x,g x')g^{\gamma'}_{\alpha'}
g^{\delta'}_{\beta'}= {\cal W}_{\alpha\beta \alpha'\beta'}(x,x')
      \, ,\end{equation}
for all $g\in SO_0(1,4)$.
\item[d)] {\bf Index symmetrizer}
\b {\cal W}_{\alpha\beta \alpha'\beta'}(x,x')={\cal W}_{\beta
\alpha \alpha' \beta'}(x,x')\, ,  \quad {\cal W}_{\alpha\beta \alpha'\beta'}(x,x')={\cal W}_{\alpha \beta
 \beta'\alpha'}(x,x')\, .\e
\item[e)] {\bf Transversality} and {\bf Tracelessness}
\begin{equation}
x^\alpha {\cal W}_{\alpha\beta
\alpha'\beta'}(x,x')=0=x'^{\alpha'}{\cal W}_{\alpha\beta
\alpha'\beta'}(x,x'), \quad {\cal W}^\alpha_{\;\;\;\alpha \alpha'\beta'} (x,x')=0={\cal
W}_{ \alpha \beta \alpha '}^{\;\;\;\;\;\;\;\;\alpha'}(x,x')\, .\end{equation}
\end{enumerate}
 The explicit knowledge of ${\cal W}$ allows us to make the quantum
field formalism work.
Using an ansatz analogous to the one used for getting the field solutions, it turns out that the tensor Wightman two-point function can be written in terms of the scalar massless minimally coupled two-point function $\cal{W}_{\textrm{mmc}}$  (thus verifying $Q_0 \cal{W}_{\textrm{mmc}}$$(x,x')=0$). 
\b {\cal W}_{\alpha\beta
\alpha'\beta'}(x,x')=\Delta_{\alpha\beta \alpha'\beta'}
(x,x'){\cal W}_{\textrm{mmc}}(x,x'), \e where
$$ \Delta(x,\partial,x',\partial')=-\frac{2}{3}{\cal S}'\theta
\theta'\cdot \left(\theta\cdot \theta'-\frac{1}{2}D_1[\theta' \cdot\bar
\partial+2H^2 \theta' \cdot x]\right)$$$$ +{\cal S}{\cal S}'\theta \cdot \theta'\left(\theta \cdot \theta'
    -\frac{1}{2}D_1[\theta' \cdot \bar \partial+2H^2
\theta' \cdot x]\right)$$ \b +\frac{1}{9}  D_2{\cal S}'\left(H^2
x\theta' \cdot+\frac{2}{3}H^2D_1\theta'Ê\cdot -\theta'\cdot \bar
\partial\right)\left(\theta\cdot\theta'
    -\frac{1}{2}D_1[\theta' \cdot\bar \partial+2H^2
\theta' \cdot x]\right)\, .\e
If one requires the function $\W_{\textrm{mmc}}$ to be dS invariant (and ignores its analyticity properties for the time being \cite{becomo,brogamo}), it will only depend on the invariant $Z(x,x')=H^2 \eta_{\alpha \beta} x^\alpha x'^{\beta}$. The equation $Q_0 \W_{\textrm{mmc}}=0$ becomes the ordinary differential equation 
\begin{equation}
(1-Z)(1+Z) \W''_{\textrm{mmc}}(Z) -8 Z \W_{\textrm{mmc}}(Z)=0\, . 
\end{equation}
In a future work \cite{gayou} we will compute the two point function for the graviton field explicitly and study its properties in detail.
 
\section{MMC in de Sitter: covariant Gupta-Bleuler \& Krein quantization}
In the previous section, we showed how to construct the tensor field $\K$ in terms of a mmc scalar field (\ref{tensorscalar}). Thus a covariant quantization of the mmc field can be of interest to  the construction  of the quantum linear gravity on dS. This field is also important in the different inflation scenarii. 
The quantization of the mmc scalar  field on dS space has been extensively studied in the literature (see \cite{becomo,chta,allen,allenfolacci,kristen,polarski} and references therein). We can summarize the results obtained as follows: a covariant construction of the propagator function for the field meets the obstacle that the Laplace-Beltrami operator $\Box_{\hat{g}}$ has a normalizable zero
mode (namely a constant mode) on the Euclidean continuation of dS space, $S^4$. Hence a dS invariant propagator inverse for the wave operator $\Box_{\hat{g}}$ does not exist. This result is not an artifact of the Euclidean continuation since Allen has proved in \cite{allen} the non-existence of a dS covariant Fock vacuum for the mmc field. Another way to understand this pathological behavior of the mmc field is to notice that the massless limit of a massive field is infrared divergent: for small $m$, the symmetric two point function develop a $\frac{1}{m^2}$ singularity. 
The wave equation of the mmc field reads as 
\begin{equation}
\label{mmceq}
Q_0 \phi(x)=0 \Longleftrightarrow \Box_{\hat{g}} \phi(x)=0\, .
\end{equation}
We work with  the following bounded global coordinates (suitable for the compactified dS space $\simeq$ Lie 
sphere $S^3\times S^1$):
\beq
\nonumber
x = (x^0 = H^{-1} \tan{\rho}, (\vec{x}, x^4) = \frac{u}{H\cos{\rho}})  \equiv (\rho, u),  \ u \in  S^3\, . 
\eeq
The coordinate $\rho, \ -\frac{\pi}{2} < \rho <\frac{\pi}{2},$ plays the role of a conformal time, whereas $\Omega$ coordinatizes the compact spacelike manifold. The set of solutions is 
\beq
\nonumber
\phi_{Llm} (x) = A_L (L e^{-i(L+2)\rho} + (L+2)e^{-iL\rho}) {\cal 
Y}_{Llm} (\Omega), \ L= 1, 2, \cdots, \ 0\leq l\leq L, \ 0\leq \vert m 
\vert \leq l\, , \eeq
where $A_L = \frac{H}{2} \lbrack 2(L+2)(L+1)L \rbrack^{-1/2}$ and the ${\cal Y}_{Llm}$ are the spherical harmonics on $S^3$. These modes form an orthonormal system with respect to the Klein-Gordon inner product, 
\beq
\label{inprod}
\langle \phi, \psi \rangle = \frac{i}{\pi^2} \int_{\rho = 0} 
\bar{\phi}(\rho, \Omega) \stackrel{\leftrightarrow}{\partial}_{\rho} 
\psi(\rho,\Omega)\, d\Omega\, . 
\eeq
The normalization constant $A_L$ breaks down at $L=0$ and this is once again the ``zero-mode'' problem. This means that one cannot obtain all the solutions of the massless wave equation (\ref{mmceq}) by taking the $m\rightarrow 0$ limit of the solutions of the massive field equation $\left(\Box_{\hat{g}}-m^2\right) \phi=0$. By solving directly the massless field equation (\ref{mmceq}), we obtain the following two solutions that replace the divergent zero mode 
\beq
\psi_g = \frac{H}{2\pi}, \ \psi_s = -\frac{iH}{2\pi}(\rho + 
\frac{1}{2} \sin{2\rho}). 
\eeq
where the constants are chosen in order to  have $\langle \psi_g, \psi_s \rangle = 1$. Searching for a covariant quantization, one must find the minimum space of solutions $\K$ (not to be confused with the graviton field in ambient space) that is dS invariant, which we find to be equal to \cite{gareta}
\beq
\K=\left\{ c_g \psi_g +\sum_{L l m} c_{L l m}\phi_{Llm}, \sum_{L l m} \left|c_{Llm}\right|^2 < \infty \right\}
\eeq
Note that the presence of the constant solution $\psi_g$ is crucial since the action of the dS group on $\phi_{Llm}$ modes generates $\psi_g$. The dS invariant space $\K$ is the space of physical states (up to gauge transformations as we will see below) \cite{gareta}. However, as an inner-product space equipped with the Klein-Gordon inner product, it is a degenerate space because the state $\psi_g$ is orthogonal to the whole space of solutions generated by the $\phi_{Llm}$ and to itself. Due to this degeneracy, canonical quantization applied to this set of modes yields once again a non covariant field. It is in particular not enough to have a representation of the dS group on the Fock space where the field operators act to guarantee the covariance of the field (see \cite{bievrerenaud} for a detailed discussion of this fact).

On the other hand, the massless theory with  Lagrangian  $\sqrt{-|\hat{g}|}\hat{g}^{\mu \nu} \partial_\mu  \phi \partial_\nu \phi$ being invariant under the transformation $\phi \rightarrow \phi +\textrm{constant}$, we are forced to deal one way or another with redundant states. In the canonical approach, a formalism \emph{\`a la} Gupta-Bleuler  \cite{ga} is well suited to deal with such gauge theories. The space of  gauge states is  the  one dimensional subspace $\cal{N}=\C$$\psi_{g}$ and the physical one particle space is precisely $\K$ up to a gauge transformation, namely the coset $\K/\cal{N}$. As we said before, the space $\K$ endowed with the Klein-Gordon inner product (\ref{inprod}) is degenerate and if one attempts to apply  canonical quantization starting from a degenerate space of solutions, then one inevitably breaks the covariance of the field. Hence we must build a non degenerate invariant space of solutions $\H$ admitting $\K$ as an invariant subspace. Together with $\cal{N}$, the latter are constituent of the so-called Gupta-Bleuler triplet $\cal{N} \subset \K \subset \H$. $\H$, the smallest complete, non-degenerate and invariant inner-product space containing $\K$ as a closed subspace is constructed in \cite{gareta} as 
\beq
\H=\H_+ \oplus \H_+^*\, , 
\eeq 
where $\H_+=\left\{ c_0 \phi_0 +\sum_{L l m\, , \, L> 0} c_{L l m}\phi_{Llm}, \sum_{L l m} \left|c_{Llm}\right|^2 < \infty \right\}$, $\phi_0$ being the Allen zero mode given by $\phi_{0} = \psi_g + \frac{1}{2}\psi_s$.
This space $\H$ is defined as a direct sum of a Hilbert space and an anti-Hilbert space (a space with definite negative inner product) which is the definition of a Krein space (see  \cite{bievrerenaud,stro} and references therein). The family $\left\{\phi_{Llm}\, , \, \phi^{\ast}_{Llm}\, , \, L>0\right\} \bigcup \{ \phi_0 \}$ is a pseudo-orthonormal basis for this Krein space, for which the non-vanishing inner products are
\beq
\left\langle \phi_{Llm,L>0}, \phi_{Llm,L>0}\right\rangle = \left\langle \phi_{0}, \phi_{0}\right\rangle=1 \quad \textrm{and} \quad 
\left\langle \phi^*_{Llm,L>0}, \phi^*_{Llm,L>0}\right\rangle = \left\langle \phi^*_{0}, \phi^*_{0}\right\rangle=-1\, .
\eeq
A crucial feature of this Krein-Gupta-Bleuler formalism is that the space $\H$ contains non physical states including negative norm states, but not restricted to them ($\psi_s \notin  \K$ and so $\phi_0 \notin  \K$ the space of physical states). Nevertheless these non physical states are indispensable as intermediate objects to assure the dS covariance and the gauge covariance of the quantum theory.   
Following \cite{min,bievrerenaud}, one can build the Fock space $\underline{\H}$ over the Krein space $\H$ and we define the quantum field $\varphi$ on $\underline{\H}$ by
\beq
\varphi = \sum_{k\geq 0}(a_k \phi_k(x) + h.c.) - \sum_{k\geq 0}(b_k  \overline{\phi}_k(x) + h.c.)\, , \ \lbrack a_k , a^{\dagger}_{k'} \rbrack 
= \delta_{kk'} = -\lbrack b_k , b^{\dagger}_{k'} \rbrack\, , 
\eeq
where k stands for $(Llm), L\geq 0$, $\phi_{k=0}$ denoting the previously defined Allen zero mode. We also define a dS invariant vacuum $\mid \Omega \,\rangle$ by 
\beq
a_k \mid \Omega \, \rangle = 0 = b_k \mid \Omega \,\rangle, \ k\geq 
0\, . 
\eeq
The whole (Krein-Fock) space $\underline{{\cal H}}$ has a 
Gupta-Bleuler structure:
\beq
\nonumber
\underline{ \underline{{\cal N}}} \subset \{ 
(a^{\dagger}_g)^{n_0}(a^{\dagger}_{k_1})^{n_1}\cdots 
(a^{\dagger}_{k_l})^{n_l}\mid \Omega \,\rangle \} \equiv 
\underline{{\cal K}} \subset \underline{{\cal H}}\, ,
\eeq
Here $\underline{ \underline{{\cal N}}}$ is the subspace of the 
physical space $\underline{{\cal K}}$ which is orthogonal to 
$\underline{{\cal K}}$. It is actually the space of gauge states.
Indeed  any physical state $\Psi \in \underline{{\cal K}}$, is equal to 
its ``gauge transform'' $\exp{-\frac{\pi\lambda}{H}(a^{\dagger}_g - 
a_g )} \Psi$ up to an element of $\underline{\underline{{\cal N}}}$.
We shall say that two such states  are physically equivalent.
An
observable $A$ is a symmetric operator on $\underline{{\cal H}}$ such 
that $\langle \Psi | A
| \Psi \, \rangle = \langle \Psi' \mid A \mid \Psi' \, \rangle$ 
for any pair of equivalent physical states.
As a matter of fact, the 
field $\varphi$ is not an observable whereas 
$\partial_{\mu}\varphi$, where $\mu$ refers to  global coordinates, is.
Therefore the stress tensor 
$$
T_{\mu \nu} = \partial_{\mu}\varphi \partial_{\nu}\varphi - 
\frac{1}{2} g_{\mu \nu} g^{\rho \sigma} \partial_{\rho}\varphi 
\partial_{\sigma}\varphi $$
is an observable. Its most remarkable feature is that it meets all 
reasonable requirements one should expect from such a physical 
quantity, namely, 
\bei
\item[(i)] No need of renormalization: 
$\vert \langle k^{n_1}_1 \cdots 
k^{n_l}_l \mid T_{\mu \nu} \mid k^{n_1}_1 \cdots k^{n_l}_l \, \rangle 
\vert < \infty$, 
\item[(ii)] Positiveness of the energy component on the 
physical sector: $\langle k^{n_1}_1 \cdots k^{n_l}_l \mid T_{00} \mid 
k^{n_1}_1 \cdots k^{n_l}_l \, \rangle \geq 0$, 
\item[(iii)] The vacuum 
energy is zero: $\langle \Omega \mid T_{00} \mid \Omega \, \rangle = 
0$.
\eni

\section{Conclusions}\label{sec4}
We sketched a dS covariant yet non pathological construction of the linear quantum gravity in dS spacetime. The difficulty to construct a reasonable dS covariant quantum gravity led naturally many authors in the past to talk about spontaneous symmetry breaking of dS spacetime.  For instance, Polyakov \cite{polya}, among others \cite{wood,antorev}, suggests that ``the cosmological constant may be screened by the infrared fluctuations of the metric, much like the electric charge in quantum electrodynamics. In the time-dependent picture the screening is equivalent to the instability of space-times with constant curvature. It is important, therefore to find out, whether the dS space carries the infrared seeds of its own destruction''.
Using our covariant formalism, we believe that some of the existing literature on the infrared effects on dS space could be clarified. One can start by computing potentially observable effects related to the spectrum of primordial gravitational wave perturbations in dS as the Sachs-Wolfe effect \cite{wolf}. One then needs to go beyond the tree level approximation in order to reexamine the possibility of a (perturbative) spontaneous breaking of dS symmetry by radiative effects \cite{coleman,mottola}. 

Finally we want to underline the fact  that Physics in dS spacetime is far from being a clear and familiar domain of investigation and that the construction of acceptable physical observables in dS is a delicate matter. For instance the usual observables of Minkowskian QFT, like scattering amplitudes, are ill defined on dS due to the existence of horizons. Aside from \cite{ilio}, very few efforts have been done in this direction. An interesting direction of investigation is the functional Schrodinger reprsentation of quantum field theory, where the existence of asymptotic states does not play any special role.

\end{document}